
\magnification
\magstep 1
\nopagenumbers
\vsize = 22 true cm
\hsize = 15 true cm
\parindent=1.2 true cm
\def\sa{\vskip 0.5 true cm}
\def\sb{\vskip 1 true cm}
\def\spar{\vskip 0.4 true cm}

\baselineskip = 0.6 true cm

\rightline{\bf LYCEN 8829}

\rightline{July 1988}

\sb
\sa

\centerline{\bf TRANSFORMATIONS GENERALIZING THE LEVI-CIVITA,}
\centerline{\bf KUSTAANHEIMO-STIEFEL, AND FOCK TRANSFORMATIONS}

\sb
\sa

\centerline{{\bf Maurice KIBLER}$^1$ and {\bf Pierre LABASTIE}$^2$}

\sb

\centerline{$^1$Institut de Physique Nucl\'eaire (et IN2P3)}

\centerline{Universit\'e Claude Bernard Lyon-1}

\centerline{43, Boulevard du 11 Novembre 1918}

\centerline{69622 Villeurbanne Cedex, France}

\sa

\centerline{$^2$LASIM (associ\'e au CNRS No. 171)}

\centerline{Universit\'e Claude Bernard Lyon-1}

\centerline{43, Boulevard du 11 Novembre 1918}

\centerline{69622 Villeurbanne Cedex, France}

\sb
\sb
\sb

\noindent Paper presented at the XVII$^{th}$ International Colloquium on
Group Theoretical Methods in Physics, Sainte-Ad\`ele (Qu\'ebec), Canada,
June 27 to July 2, 1988. Published in {\it Group Theoretical Methods in
Physics}, eds. Y. Saint-Aubin and L. Vinet (World Scientific, Singapore,
1989). p.~660.

\vfill\eject

\baselineskip = 0.60 true cm

\centerline{\bf TRANSFORMATIONS GENERALIZING THE LEVI-CIVITA,}
\centerline{\bf KUSTAANHEIMO-STIEFEL, AND FOCK TRANSFORMATIONS}

\sb
\sb

\baselineskip = 0.44 true cm

\centerline{Maurice {\bf KIBLER}$^1$ and Pierre {\bf LABASTIE}$^2$}

\vskip 0.18 true cm

\centerline{$^1$Institut de Physique Nucl\'eaire (et IN2P3)}

\centerline{$^2$LASIM (associ\'e au CNRS No. 171)}

\centerline{Universit\'e Claude Bernard Lyon-1}

\centerline{69622 Villeurbanne Cedex, France}

\sb
\sa

\centerline{ABSTRACT}

\vskip 0.28 true cm
\baselineskip = 0.46 true cm

\leftskip = 1.5 true cm
\rightskip = 1.5 true cm

\noindent Preliminary results concerning non-quadratic (and non-bijective)
transformations that exibit a degree of parentage with the well known
Levi-Civita, Kustaanheimo-Stiefel, and Fock
transformations are reported in this article. Some of the new transformations
are applied to non-relativistic quantum dynamical systems in two dimensions.

\leftskip = 0 true cm
\rightskip = 0 true cm

\baselineskip = 0.59 true cm

\sa
\sb

\noindent {\bf 1. Introduction and Preliminaries}

\vskip 0.23 true cm

Non-bijective (canonical) transformations have received a great deal of
attention in the recent years. In particular, {\bf quadratic} transformations
generalizing the so-called Levi-Civita$^{1)}$ and Kustaanheimo-Stiefel$^{2)}$
transformations have been studied from algebraic, geometrical and
Lie-like viewpoints.$^{3-12)}$

It is the aim of the present paper to
extend the study in Ref. 9 to {\bf
non-quadratic} transformations. We shall see that the Fock$^{13)}$
(stereographic) transformation belongs to the set of the transformations
introduced in this work.

Following the work of Lambert and Kibler$^{9)}$ on quadratic transformations,
we shall use Cayley-Dickson algebras as a general framework
for defining non-quadratic
transformations.  Indeed, we shall restrict ourselves to
$2m$-dimensional Cayley-Dickson algebras
$A(c) \equiv A(c_1, c_2, \ldots, c_p)$,
$c_i = \pm 1$ with $i = 1, 2, \ldots, p$, for which $2^p$ $($$=$ $2m) \le 8$.
The cases $2m = 2$, 4, and 8 correspond to $A(c_1) = C$ or $\Omega$,
$A(c_1,c_2) = H$ or $N_1$, and $A(c_1,c_2,c_3) = O$ or $O^{\prime}$, i.e.,
to the
algebras of complex numbers or hyperbolic complex numbers, quaternions or
hyperbolic quaternions, and octonions or hyperbolic octonions, respectively.

A basic ingredient for generating non-bijective transformations depends
on the fact that the product $x = uv$ of two hypercomplex numbers $u$ and $v$
in $A(c)$ can be written in a matrix form as
${\bf x} = A({\bf u}) {\bf v}$,
where $A({\bf u})$ is a $2m \times 2m$ matrix generalizing the Hurwitz
matrix.$^{9)}$ An important property of $A({\bf u})$ for what follows is
$\tilde A({\bf u})^N \eta A({\bf u})^N = (\tilde {\bf u} \eta {\bf u})^N \eta$
for $N \in Z$, where the metric $\eta$ reads
$\eta = {\rm diag}(1,-c_1,-c_2,c_1c_2,-c_3,c_1c_3,c_2c_3,-c_1c_2c_3)$ in
the case $2m = 8$.
(The cases $2m = 4$ and $2$ may be deduced from the case $2m = 8$ by simple
dimensional reduction.)

Another
ingredient is provided by the anti-involutions described at length in
Ref. 9. Let us recall that, in a $2m$-dimensional Cayley-Dickson algebra
$A(c)$, it is possible to construct
$2m - \delta(m,1)$ anti-involutions including the complex conjugation.
We shall use $j$
to denote such anti-involutions. In matrix form, the product $x = uj(v)$ of
the two hypercomplex numbers $u$ and $j(v)$ in $A(c)$ is given by
${\bf x} = A({\bf u}) \epsilon {\bf v}$,
where $\epsilon$ is a $2m \times 2m$ diagonal matrix associated with the
anti-involution $j$ of $A(c)$.

We are now in a position to define
transformations that we shall classify
according to the nomenclature $A_N$, $B_N$, and $C_N$. These will be
defined in sections 2, 3, and 4, respectively. Section
5 will deal with applications of some of the
new transformations to various $R^2$ potentials. Work on
transformations in higher
dimensions and on applications to other potentials (e.g., the H\'enon-Heiles
potential) is in progress.
\spar

\noindent {\bf 2. Transformations of Type $A_N$}

\vskip 0.23 true cm

The map $A(c) \to A(c)$ : $u \mapsto x = u^{N+1}$ with $N \in Z$ gives rise to
a $R^{2m} \to R^{2m}$ transformation defined through
$${\bf x} = A({\bf u})^N {\bf u} \eqno(1)$$
This is a transformation of magnitude $N+1$ since
$r^2 = \rho^{2(N+1)}$,
where $r^2 = \tilde {\bf x} \eta {\bf x}$ and $\rho^2 = \tilde {\bf u} \eta
{\bf u}$.

We shall refer to the transformations (1) as transformations of type $A_N$. The
transformations of type $A_1$ (or quasiHurwitz transformations) have been
investigated in Ref. 9. They correspond to fibrations on
spheres ($S^{2m-1} \to S^{2m-1}/Z_2$) or to fibrations on hyperboloids
($H^{2m-1}(m,m) \to H^{2m-1}(m,m)/Z_2$). As a typical example of a
transformation of type $A_1$, for $2m = 2$,
$c_1 = -1$, and $N=1$, we obtain the Levi-Civita ($R^2 \to R^2$)
transformation associated with the fibration $S^1 \to S^1/Z_2 = RP^1$.

Still for $N=1$, the line element
$ds^2 = d \tilde {\bf x} \> \eta \> d {\bf x}$ has been given in Ref. 9 for
an arbitrary transformation of type $A_1$. As a consequence, the
Laplace-Beltrami operator may be obtained in a straightforward way for such a
transformation. By way of illustration,
for the transformation introduced in Ref. 5 and
corresponding to
$2m = 4$ and $c_1 = c_2 = -1$, we get
$$\Delta_x = (1/4 \rho ^2) \{ \Delta_u$$
$$+ (1/u_1^2) [ (u_2^2 + u_3^2) \partial_{u_4u_4}
             +(u_3^2 + u_4^2) \partial_{u_2u_2}
             +(u_4^2 + u_2^2) \partial_{u_3u_3} ]$$
$$- (2/u_1^2) (u_2u_3 \partial_{u_2u_3}
             + u_3u_4 \partial_{u_3u_4}
             + u_4u_2 \partial_{u_4u_2})$$
$$+ (2/u_1^2) ( u_1 \partial_{u_1}
               -u_2 \partial_{u_2}
               -u_3 \partial_{u_3}
               -u_4 \partial_{u_4} ) \} $$
$$\Delta_x = \sum_{\alpha = 1}^4 \partial_{x_{\alpha}x_{\alpha}}, \quad
      \Delta_u = \sum_{\alpha = 1}^4 \partial_{u_{\alpha}u_{\alpha}}, \quad
\rho ^2 = \sum_{\alpha = 1}^4 u_{\alpha}^2 \eqno(2)$$

General properties of the transformations of type $A_N$ with $N \in Z -
\{-1\}$ may be easily derived in the case $2m = 2$. We have the
5 following properties
$$x_1^2 - c_1 x_2^2 = (u_1^2 - c_1 u_2^2)^{N+1}$$
$$d{\bf x} = (N+1) A({\bf u})^N d{\bf u}$$
$$\vec {\nabla}_{\bf x}
= (N+1)^{-1}(u_1^2-c_1u_2^2)^{-N}\eta A({\bf u})^N\eta \vec {\nabla}_{\bf u}$$
$$dx_1^2 - c_1 dx_2^2 = (N+1)^2(u_1^2-c_1u_2^2)^N(du_1^2-c_1du_2^2)$$
$$\partial_{x_1x_1} - c_1 \partial_{x_2x_2}
= (N+1)^{-2}(u_1^2 - c_1 u_2^2)^{-N}(\partial_{u_1u_1}-c_1\partial_{u_2u_2})
\eqno(3)$$

The situation where $N = -1$ deserves a special treatment. Indeed, the
transformations of type $A_{-1}$ are not interesting since they correspond to
$A(c) \to A(c)$ : $ u \mapsto x = 1$. In the particular
case $2m = 2$, we define
transformations of the type $A'_{-1}$ in the
following way. We start with the column
vector ${\bf \omega} = A({\bf u})^{-1} d{\bf u}$. The elements
of ${\bf \omega}$ are
one-forms which turn out to be complete
differentials. A direct integration leads
to the $R^2 \to R^2$ transformations $(u_1, u_2) \mapsto (x_1, x_2)$ with
$$x_1 = (1/2) \> \ln (u_1^2 - c_1 u_2^2)
\quad {\rm for}  \quad u_1^2 - c_1 u_2^2 > 0$$
$$\tan x_2 = u_2/u_1 \ {\rm for} \ c_1 = -1 \quad {\rm or}
\quad \tanh x_2 = u_2/u_1 \ {\rm for} \ c_1 = +1 \eqno(4)$$
The latter transformations, referred to as transformations of type $A'_{-1}$,
correspond to the map $A(c_1) \to A(c_1)$ : $u \mapsto x = \log u$. These
transformations satisfy the properties
$$\vec {\nabla}_{\bf x} = \tilde A({\bf u}) \vec {\nabla}_{\bf u}$$
$$dx_1^2 - c_1 dx_2^2=(u_1^2 - c_1 u_2^2)^{-1}(du_1^2 - c_1 du_2^2)$$
$$\partial_{x_1x_1} - c_1 \partial_{x_2x_2}
= (u_1^2 - c_1 u_2^2)(\partial_{u_1u_1} - c_1 \partial_{u_2u_2}) \eqno(5)$$
\spar

\noindent {\bf 3. Transformations of Type $B_N$}

\vskip 0.23 true cm

We now write Eq. (1) in a slightly modified form: the relation
$${\bf x} = A({\bf u})^N \epsilon {\bf u}, \qquad N \in Z \eqno(6)$$
defines a $R^{2m} \to R^{2m - n}$ transformation
($n \ge 0$) that we shall call a
transformation of type $B_N$. This transformation
(of magnitude $N + 1$) corresponds to the
map $A(c) \to A(c)$ : $u \mapsto x = u^N j(u)$.

As a first typical example, we take $2m = 4$, $c_1 = c_2 = -1$, $N = 1$, and
$\epsilon = {\rm diag}(1,1,1,-1)$. Then,
we obtain a $R^4 \to R^3$ surjection that
is nothing but the Kustaanheimo-Stiefel transformation (associated with
the Hopf
fibration $S^3 \to S^2$ of compact fiber $S^1$). More generally, the
$R^{2m} \to R^{2m-n}$ transformations of type $B_1$
(or Hurwitz transformations) correspond to $n = 2m-1$ or
$n = m - 1 + \delta(m,1)$. They have been studied by
Lambert and Kibler.$^{9)}$ For {2m} fixed and $n = m - 1 + \delta(m,1)$,
there are 3 classes of transformations of
type $B_1$ associated with (i) Hopf fibrations on spheres
($S^{2m-1} \to S^{m - \delta(m,1)}$ with compact fiber
$S^{m-1 + \delta(m,1)}$), (ii) fibrations on hyperboloids with compact fibers,
and (iii) fibrations on hyperboloids with non-compact fibers.

As a second typical example, let us set $2m = 4$, $N = -1$, and
$\epsilon = {\rm diag}(1,1,1,-1)$. Thus, Eq. (6) yields the $R^4 \to R^4$
transformation
$$x_1 = {1\over \rho^2} (u_1^2 - c_1  u_2^2 - c_2 u_3^2 - c_1 c_2 u_4^2)$$
$$x_2 = - {2\over \rho^2} c_2 u_3 u_4, \quad
  x_3 = {2\over \rho^2} c_1 u_2 u_4, \quad
  x_4 = - {2\over \rho^2} u_1 u_4 \eqno(7)$$
where $\rho^2 = u_1^2 - c_1 u_2^2 - c_2  u_3^2 + c_1 c_2 u_4^2$.
For $c_1 = c_2 = -1$, the
latter transformation particularizes to the $R^4 \to S^3$ stereographic
projection known as the Fock$^{13)}$ projection. Similar results may be
obtained in the cases $2m = 2$ and 8. In particular, for $2m = 8$,
$c_1 = c_2 = c_3 = -1$, $N = -1$, and
$\epsilon = \epsilon_k$ ($k = 1, 2, \cdots, 7$), see Ref. 9,
we get a $R^8 \to S^7$ projection.
\spar

\noindent {\bf 4. Transformations of Type $C_N$}

\vskip 0.23 true cm

A last class of transformations is obtained when the matrix $\epsilon$ in Eq.
(6) is replaced by an (arbitrary) matrix which is neither the identity matrix
(yielding transformations of type $A_N$) nor a matrix associated with an
anti-involution $j$ of $A(c)$ (yielding transformations of type $B_N$). Some
transformations of type $C_1$ (or pseudoHurwitz transformations) have been
described elsewhere.$^{9,10)}$
In the case
$2m = 8$, the
transformations of type $C_1$ corresponding to diagonal matrices $\epsilon$
having $\pm 1$ for matrix elements have been classified in Ref. 9.

\spar

\noindent {\bf 5. Applications}

\vskip 0.23 true cm

First, let us consider the two-dimensional Schr\"odinger equation
$$( - {1\over 2} \Delta_x - {Z\over r^{\alpha}})\psi
= E\psi, \qquad \alpha \in R \eqno(8)$$
where $r = (x_1^2 + x_2^2)^{1/2}$. The application to Eq. (8)
of a ($R^2 \to R^2$)
transformation of type $A_N$ with $2m = 2$, $c_1 = -1$, and $N \in Z - \{-1\}$
leads to the partial differential equation
$$[- {1\over 2} \Delta_u - (N+1)^2 E \rho^{2N}] \hat \psi =
(N+1)^2 Z \rho^{2N - \alpha (N+1)} \hat \psi \eqno(9)$$
where $\rho = (u_1^2 + u_2^2)^{1/2}$ and $\hat \psi \equiv \hat \psi (u)$ is
the transform of $\psi \equiv \psi (x)$ under the considered transformation.
Furthermore, let us impose that $2N - \alpha (N + 1) = 0$. Then, the
transformation of type $A_N$
allows to transform the $R^2$ Schr\"odinger equation for
the potential $-Z(x_1^2 + x_2^2)^{-N/(N+1)}$ and the energy $E$ into
the $R^2$ Schr\"odinger equation for
the potential $-(N + 1)^2E(u_1^2 + u_2^2)^N$ and the energy $(N+1)^2Z$. (Note
that in such a transformation the roles of the energy $E$ and the coupling
constant $Z$ are interchanged.) The solutions for
$\alpha$ ($=$ $2N/(N+1)$) $\in Z$ and $N \in Z -\{-1\}$ correspond to
$(\alpha, N) = (1,1)$, $(3,-3)$, and $(4,-2)$. (The solution $(0,0)$ is
trivial!) In other words, the Schr\"odinger equations for the potentials
$1/r$ (Coulomb), $1/r^3$, and $1/r^4$ are transformed into Schr\"odinger
equations for the potentials $\rho^2$ (harmonic oscillator), $1/\rho^6$, and
$1/\rho^4$, respectively.

Second, we consider the Schr\"odinger equation
$$[ - {1\over 2} \Delta_u - {Z\over {(u_1^2 + u_2^2)^{1/2}}}] \hat \psi =
E \hat\psi \eqno(10)$$
for a two-dimensional hydrogen atom. By using the transformation of type
$A'_{-1}$ with $c_1 = -1$, Eq. (10) may be converted into
$$( - {1\over 2} \Delta_x - Z {\rm e}^{x_1} - E {\rm e}^{2x_1}) \psi
= 0 \eqno(11)$$
Since $x_2$ is a cyclical coordinate, we can set
$-(1/2) \partial_{x_2x_2} \psi = -K \psi$ so that we arrive at
$$( - {1\over 2} \partial_{x_1x_1} - Z {\rm e}^{x_1} - E {\rm e}^{2x_1})
\psi = K \psi \eqno(12)$$
which may be recognized as the Schr\"odinger equation for an one-dimensional
Morse potential (provided $Z > 0$ and $E < 0$). (The usual Morse$^{14)}$
potential is reverted in the $x_1$ variable.)

Third, we close with the $R^2$ Schr\"odinger equation
$$[ - {1\over 2} \Delta_u + V_0 \ln (u_1^2 + u_2^2)] \hat \psi
= E \hat \psi \eqno(13)$$
The transformation of type $A'_{-1}$ with $c_1 = -1$ makes
it possible to change Eq. (13) into
$$( - {1\over 2} \Delta_x - E {\rm e}^{2x_1} + 2 V_0 x_1 {\rm e}^{2x_1} )
\psi = 0 \eqno(14)$$
and the separation of variables
($-(1/2) \partial_{x_2x_2} \psi = -K \psi$)
leads to
$$[ - {1\over2} \partial_{x_1x_1} - (E - 2 V_0 x_1) {\rm e}^{2x_1}] \psi =
K \psi \eqno(15)$$

\spar

\noindent {\bf References}

\vskip 0.23 true cm
\parindent = 0.9 true cm
\baselineskip = 0.53 true cm

\item{[1]} Levi-Civita, T., cited in Ref. 2.

\item{[2]} Kustaanheimo, P. and Stiefel, E.,
J. reine angew. Math. {\bf 218}, 204 (1965).

\item{[3]} Boiteux, M., Physica {\bf 75}, 603 (1974); J. Math. Phys. {\bf 23},
1311 (1982).

\item{[4]} Polubarinov, I.V., ``On Application of Hopf Fiber Bundles in
Quantum Theory'', preprint E2-84-607, JINR: Dubna (1984).

\item{[5]} Kibler, M. and N\'egadi, T., Croat. Chem.
Acta, CCACAA, {\bf 57}, 1509 (1984).

\item{[6]} Iwai, T., J. Math. Phys. {\bf 26}, 885 (1985).

\item{[7]} Lambert, D., Kibler, M., and Ronveaux, A., in {\it Proc. 14th Int.
Coll.
Group Theoretical Methods in Physics}, ed. Y.M. Cho, World Scientific:
Singapore (1986). (p 304)

\item{[8]} Lambert, D. and Kibler, M., in {\it Proc. 15th Int. Coll.
Group Theoretical Methods in Physics}, ed. R. Gilmore, World Scientific:
Singapore (1987). (p 475)

\item{[9]} Lambert, D. and Kibler, M., J. Phys. A: Math. Gen. {\bf 21},
307 (1988).

\item{[10]} Kibler, M., ``Transformations canoniques non bijectives'',
Report LYCEN 8713, IPN: Lyon (1987). (unpublished)

\item{[11]} Kibler, M. and Winternitz, P., J. Phys. A: Math. Gen. {\bf 21},
1787 (1988).

\item{[12]} Kibler, M., in {\it Proc. 16th Int. Coll.
Group Theoretical Methods in Physics}, eds. H.D. Doebner and T.D. Palev,
Springer: Berlin (1988).

\item{[13]} Fock, V.A., Z. Phys. {\bf 98}, 145 (1935).

\item{[14]} Morse, P.M., Phys. Rev. {\bf 34}, 57 (1929).
\bye